\documentclass[preprint,showpacs,showkeys,preprintnumbers,
amsmath,amssymb,aps,floatfix,pre]{revtex4}

\usepackage{graphicx}
\usepackage{multirow}
\usepackage{times}
\usepackage{amssymb}
\usepackage{dcolumn}%
\usepackage{bm}%
\usepackage{hyperref}%
\usepackage{upgreek}

\hypersetup
{
	colorlinks=true,
	linkcolor=blue,
	citecolor=blue
}

\begin{document}

\title[]{Cage rattling does not correlate with the local geometry in molecular liquids}
\author{S.\@ Bernini}
\affiliation{Dipartimento di Fisica ``Enrico Fermi'', 
Universit\`a di Pisa, Largo B.\@Pontecorvo 3, I-56127 Pisa, Italy}
\author{F.\@ Puosi}
\affiliation{Laboratoire Interdisciplinaire de Physique, Universit\`e Joseph Fourier Grenoble, CNRS,	
 Ê38402 Saint Martin d'H\`eres, France}
\author{D.\@ Leporini}
\email{dino.leporini@df.unipi.it}
\affiliation{Dipartimento di Fisica ``Enrico Fermi'', 
Universit\`a di Pisa, Largo B.\@Pontecorvo 3, I-56127 Pisa, Italy}
\affiliation{IPCF-CNR, UOS Pisa, Italy}

\date{\today}% It is always \today, today,

\begin{abstract}
Molecular-dynamics simulations of a liquid of short linear molecules have been 
performed to investigate the correlation between the particle dynamics in the 
cage of the neighbors and the local geometry. The latter is characterized in 
terms of the size and the asphericity of the Voronoi polyhedra. The correlation 
is found to be poor. In particular,  in spite of the different Voronoi volume 
around the end and the inner monomers of a molecule, all the monomers exhibit 
coinciding displacement distribution when they are caged (as well as at longer 
times during the structural relaxation). It is concluded that the fast dynamics 
during the cage trapping is a non-local collective process involving monomers 
beyond the nearest neighbours.
\end{abstract}

%Use showkeys class option if keyword
\maketitle

\section{Introduction}

Global order is virtually absent in systems like glasses and liquids. On the 
contrary, local order is present in both disordered and ordered phases  
\cite{FrenkelPRL95,TorquatoStilliPRE02,AsteJPCM05,LocalOrderJCP13}. A crucial 
aspect of the solidification leading to a glass is that it is associated only to 
subtle static structure changes, e.g. the  static structure factor $S(q)$, 
measuring the spatial correlations of the particle positions,  does not show 
appreciable changes on approaching the glass transition (GT). The absence of 
apparent static correlations distinguishing a glass and a liquid motivated 
research where the focus is on the dynamical facilitation by mobile particles  
on the nearby ones to trigger their kinetics \cite{ChandlerGarra10}. The 
facilitation approach is seen as an attempt to put in a more microscopic way the 
concept of free volume by  assuming that  geometrical constraints act at the 
level of kinetic rules with no reference to the microscopic interactions which 
are responsible for them \cite{BerthierBiroliRMP11}.

A different line of thought suggests that structural aspects matter in the  
dynamical behaviour of glassforming systems. This includes the Adam-Gibbs 
derivation of the structural relaxation \cite{AdamGibbs65,DudowiczEtAl08} - 
built on the thermodynamic notion of the configurational entropy 
\cite{GibbsDiMarzio58} -, the mode- coupling theory \cite{GotzeBook} and 
extensions \cite{SchweizerAnnRev10}, the random first-order transition theory 
(RFOT) \cite{WolynesRFOT07}, the frustration-based approach \cite{TarjusJPCM05}, 
as well as the so-called  elastic models  \cite{Dyre06,Puosi12}.  The search of 
a link between structural ordering and slow dynamics motivated several studies 
in liquids 
\cite{NapolitanoNatCom12,EdigerDePabloNatMat13,RoyallPRL12,BarbieriGoriniPRE04}, 
colloids \cite{StarrWeitz05,TanakaNatMater08,TanakaNatCom12} and polymeric 
systems 
\cite{StarrWeitz05,DePabloJCP05,GlotzerPRE07,LasoJCP09,BaschnagelEPJE11,
MakotoMM11,LariniCrystJPCM05}. 

On approaching the glass transition, particles are trapped by the cage of  the 
first neighbors more effectively and the average escape time, i.e. the 
structural relaxation time $\tau_\alpha$, increases from a few picoseconds up to 
thousands of seconds  
\cite{EdigerHarrowell12,AngelNgai00,DebeStilli2001,Richert02,BerthierBiroliRMP11
}.  The caged particles are not completely immobilized by the surroundings but 
they wiggle with mean-square amplitude $\langle u^2\rangle$  on the picosecond 
time scale $t^\star$. $\langle u^2\rangle$ is related to the Debye-Waller factor 
which, assuming harmonicity of thermal motion, takes the form $\exp\left( -q^2 
\langle u^2\rangle / 3 \right)$ where $q$ is the absolute value of the 
scattering vector. Henceforth, $\langle u^2\rangle$ will be referred to as 
short-time mean-square displacement (ST-MSD). The temporary trapping and 
subsequent escape mechanisms lead to large fluctuations around the averaged 
dynamical behavior with strong heterogeneous dynamics \cite{BerthierBiroliRMP11} 
and leads to non-exponential relaxation and aging \cite{MonthBouch96}.
Despite the huge range of time scales older \cite{TobolskyEtAl43} and recent  
theoretical 
\cite{Angell95,HallWoly87,Dyre96,MarAngell01,Ngai04,Ngai00,
DouglasCiceroneSoftMatter12} studies addressed the  rattling process in the cage 
to understand the structural relaxation - the escape process -  gaining support 
from numerical 
\cite{Angell68,Nemilov68,Angell95B,StarrEtAl02,BordatNgai04,Harrowell06,
Harrowell_NP08,HarrowellJCP09,DouglasEtAlPNAS2009,XiaWolynes00,DudowiczEtAl08,
DouglasCiceroneSoftMatter12,OurNatPhys,lepoJCP09,Puosi11,SpecialIssueJCP13,
UnivSoftMatter11,CommentSoftMatter13,OttochianLepoJNCS11,UnivPhilMag11,Puosi12SE
,Puosi12,PuosiLepoJCPCor12,PuosiLepoJCPCor12_Erratum} and experimental works on 
glassforming liquids 
\cite{BuchZorn92,AndreozziEtAl98,DouglasCiceroneSoftMatter12} and glasses 
\cite{MarAngell01,ScopignoEtAl03,SokolPRL,Buchenau04,NoviSoko04,NovikovEtAl05,
Johari06}.

Recently,  extensive molecular-dynamics (MD) simulations evidencing the  
universal correlation between the structural relaxation time $\tau_\alpha$ and 
$\langle u^2\rangle$  were reported in polymeric systems 
\cite{OurNatPhys,lepoJCP09,Puosi11}, binary atomic mixtures 
\cite{SpecialIssueJCP13}, colloidal gels \cite{UnivSoftMatter11} and 
antiplasticized polymers \cite{DouglasCiceroneSoftMatter12}
and compared with the experimental data concerning several glassformers in  a  
wide fragility range ($20 \le m \le 191$) 
\cite{OurNatPhys,UnivPhilMag11,SpecialIssueJCP13,CommentSoftMatter13}.
One major finding was that states with equal ST-MSD $\langle u^2\rangle$ have equal relaxation times $\tau_\alpha$ too. 
For polymers states with equal ST-MSD show also equal chain reorientation rate  
\cite{lepoJCP09,Puosi11} and diffusivity \cite{Puosi11}. Diffusion scaling was 
also observed in atomic mixtures \cite{SpecialIssueJCP13}. More recently, the 
influence of free volume and the proper time scales to observe the genuine fast 
dynamics have been considered \cite{OttochianLepoJNCS11,UnivPhilMag11} as well 
as the breakdown of the Stokes-Einstein (SE) law \cite{Puosi12SE}, the relation 
with the elastic modulus \cite{Puosi12} and the spatial extension of the 
involved particle displacements at short-times  
\cite{PuosiLepoJCPCor12,PuosiLepoJCPCor12_Erratum}. 

Insight into the scaling between relaxation and fast dynamics is provided by  
the particle displacement distribution, i.e. the incoherent, or self part, of 
the van Hove function $G_{s}(r,t)$  \cite{Egelstaff:1992fk,HansenMcDonaldIIIEd}. 
The interpretation of $G_{s}(r,t)$ is direct. The product $G_{s}(r,t) \cdot 4\pi 
r^{2} \, dr$ is the probability that the particle is at a distance between $r$ 
and $r+dr$ from  the initial position after a time $t$. In terms of  $G_{s}({ 
r},t)$ the scaling property is expressed by stating that, if two physical 
states, say X and Y, are characterized by the {\it same} displacement 
distribution $G_{s}( r,t^\star)$ at the rattling time $t^\star$, they also 
exhibit the {\it same} distribution at long times, e.g. at $\tau_\alpha$ 
\cite{Puosi11,SpecialIssueJCP13}:
 \begin{equation}
G_{s}^{(X)}({ r},t^\star) = G_{s}^{(Y)}({ r},t^\star)    \iff G_{s}^{(X)}({ 
r},\tau_\alpha) = G_{s}^{(Y)}({r},\tau_\alpha)
\label{vanhovescaling}
\end{equation}
Eq.\ref{vanhovescaling} holds even in the presence of very strong dynamical  
heterogeneity  where both diffusive and jump-like dynamics are observed 
\cite{Puosi11} and,  in this respect, is consistent with previous conclusions 
that the long-time dynamical heterogeneity is predicted by the fast 
heterogeneities \cite{Harrowell06,note3}.

The present paper aims at investigating by molecular-dynamics (MD) simulations   
if the cage dynamics on the fast time scale $t^\star$ correlates with the local 
order in a molecular liquid. Local order will be characterized in terms of two 
measures, i.e. the  volume and the asphericity of the Voronoi polyhedron (VP) 
surrounding a tagged particle. VP asphericity has been analyzed also in water  
\cite{JedloJCP,Wikfeldt10,Stirnemann12}, small molecules \cite{H2SJCP}, hard 
spheres \cite{Krekelberg06} and polymers  
\cite{SegaJCP,StarrEtAl02,DouglasCiceroneSoftMatter12,Rigbyt}. For a VP with  
$V_v$ volume and $A_v$ surface the asphericity is defined as: 
\begin{equation}
 a_v=\frac{\left(A_v\right)^3}{36\pi \left(V_v\right)^2}-1
 \label{aspher}
\end{equation}
The asphericity vanishes for spheres, and is positive for not spherical objects. 
 It will be shown that the correlation between ST-MSD and the two measures of 
the local order is quite poor pointing to the non-local character of the fast 
motion of a particle within the cage of its neighbors.

The paper is organized as follows. In Sec. \ref{numerical} the molecular
model and the MD algorithms are presented. The results are
discussed in Sec. \ref{resultsdiscussion}. Finally, the main conclusions
are summarized in Sec. \ref{conclusions}.

\section{Methods}
\label{numerical}

Molecular-dynamics (MD) simulations of a melt of
 fully-flexible linear chains are performed.
The interacting potential between non-bonded monomers 
reads as:

\begin{equation}
U_{p,q}(r)=\frac{\varepsilon}{q-p} \left[ p \left( \frac{\sigma^*}{r}\right)^q
-q\left(\frac{\sigma^*}{r}\right)^p \right] + U_{cut} 
\end{equation}

Changing the $p$ and $q$ parameters does not affect 
the position $r = \sigma^* = \sqrt[6]{2}\,\sigma$ and the depth
of the potential minimum $\varepsilon$. The constant $U_{cut}$
is chosen to ensure $U_{p,q}(r) = 0 $ at $r \geq 2.5 \, \sigma$. Notice that  
$p=6,q=12$ yields the familiar Lennard-Jones (LJ) potential.
The bonded monomers interact by a potential which is the sum of the LJ potential 
 and the FENE (finitely extended
nonlinear elastic) potential \cite{GrestPRA33}:
\begin{equation}
 U_{FENE}(r)=-\frac{1}{2}kR_0^2\ln\left(1-\frac{r^2}{R_0^2}\right)
\end{equation}
where $k$ measures the magnitude of the interaction and $R_0$ is
the maximum elongation distance. The parameters $k$ and $R_0$
have been set  to $30 \, \varepsilon  / \sigma^2 $ and $ 1.5\,\sigma $
respectively. We study systems of about $2000$ monomers at different density  
$\rho$,
temperature $T$ and $p,q$ parameters. 
Each state is labeled by
the multiplet $\{\rho,p,q,T\} $.
All quantities are in reduced units: length in
units of $\sigma$, temperature in units of $\varepsilon/k_B$ 
and time in units of $\sigma \sqrt{\mu / \varepsilon}$ where 
$\mu$ is the monomer mass. We set $\mu = k_B = 1$.
$NPT$ and $NVT$ ensembles have been used for equilibration
runs, while $NVE$ ensemble has been used for production runs
for a given state point ($NPT$: constant number of particles, pres-
sure and temperature; $NVT$: constant number of particles,
volume and temperature; $NVE$: constant number of particles,
volume and energy). $NPT$ and $NVT$ ensembles are studied 
by the extended system method introduced by Andersen \cite{Andersen80} and
Nos\'e \cite{NTVnose}. The numerical integration of the augmented 
Hamiltonian is performed through the multiple time
steps algorithm, reversible Reference System Propagator Algorithm (r-RESPA)
\cite{respa}. We investigate three sets of states of a melt of trimers ($M=3$):
\begin{itemize}
 \item Set A: (1.015,7,15,1.05), (1.09,8,12,1.4)
 \item Set B: (1.016,7,15,0.7), (1.086,6,12,0.7)
 \item Set C: (0.984,6,12,0.33), (1.086,6,12,0.63)
\end{itemize}
The states of each set have equal  ST-MSD $\langle u^2\rangle$ and structural  
relaxation time $\tau_\alpha$ (see Sec.\ref{dynamic}). In addition we also study 
a melt of decamers ($M=10$) with $\rho = 1.086$, $T=0.65$ with LJ interaction 
between non-bonded monomers exhibiting a structural relaxation time $\tau_\alpha 
\simeq 152$.

\begin{figure}[t]
\begin{center}
\includegraphics[width=0.5\linewidth]{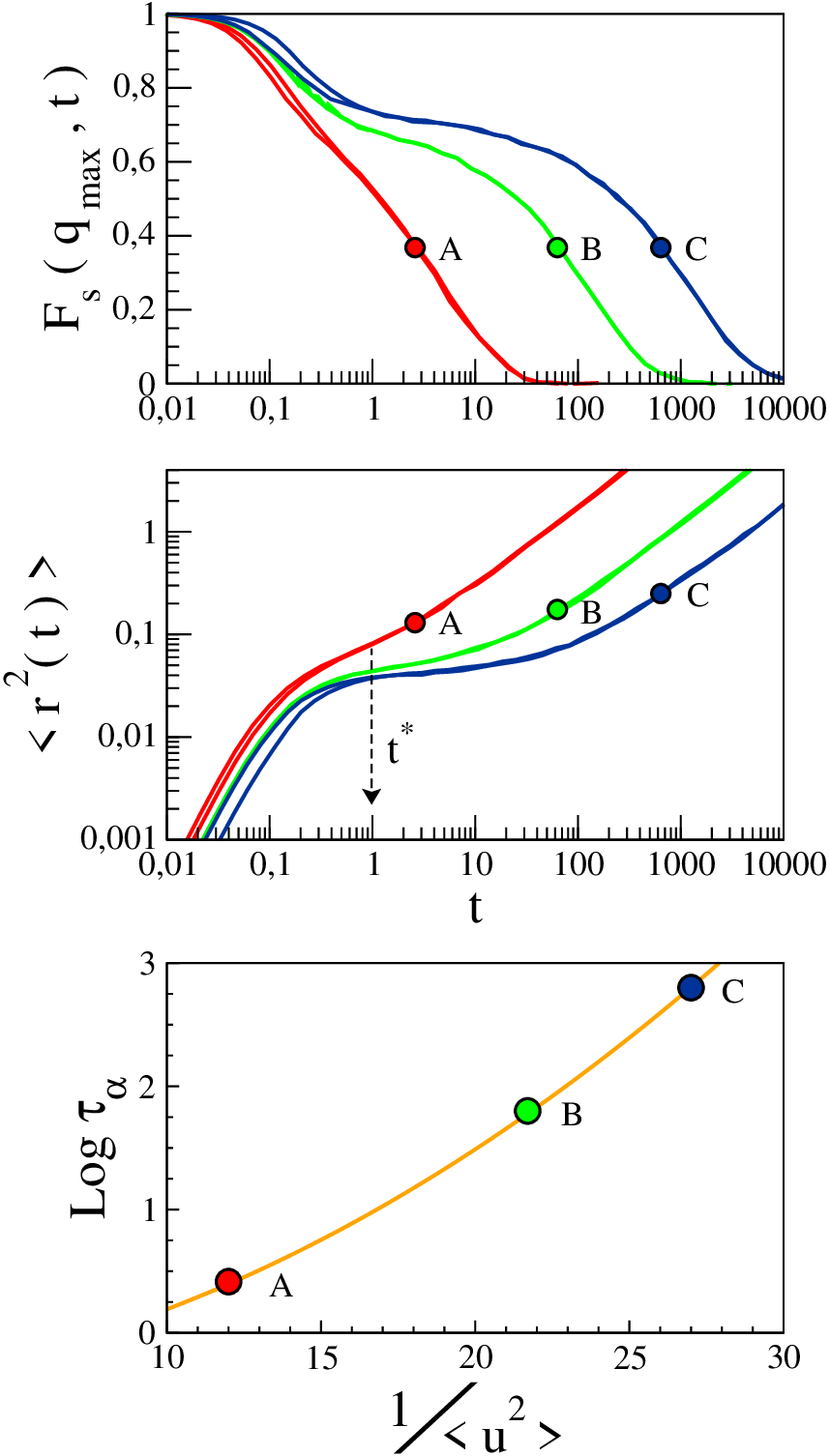}
\end{center}
\caption{Top: ISF of the states of the sets A,B,C; the dots mark the
structural relaxation time $\uptau_\alpha$. Middle: monomer MSD of the states  
of the sets A,B,C. $\langle u^2 \rangle$ is the short-time MSD (ST-MSD) 
evaluated at the time
$t^*$ when MSD changes the concavity in the log-log plot.  Bottom: location of  
the sets of states in the master curve $\log \tau_\alpha$ vs $\langle u^2 
\rangle^{-1}$ (orange line) \cite{OurNatPhys}. Note that ISF and MSD of states 
of the same set coincide from $ t \simeq t^{\star}$ onward.
}
 \label{fig1}
\end{figure}

\begin{figure}[t]
\begin{center}
\includegraphics[width=0.5\linewidth]{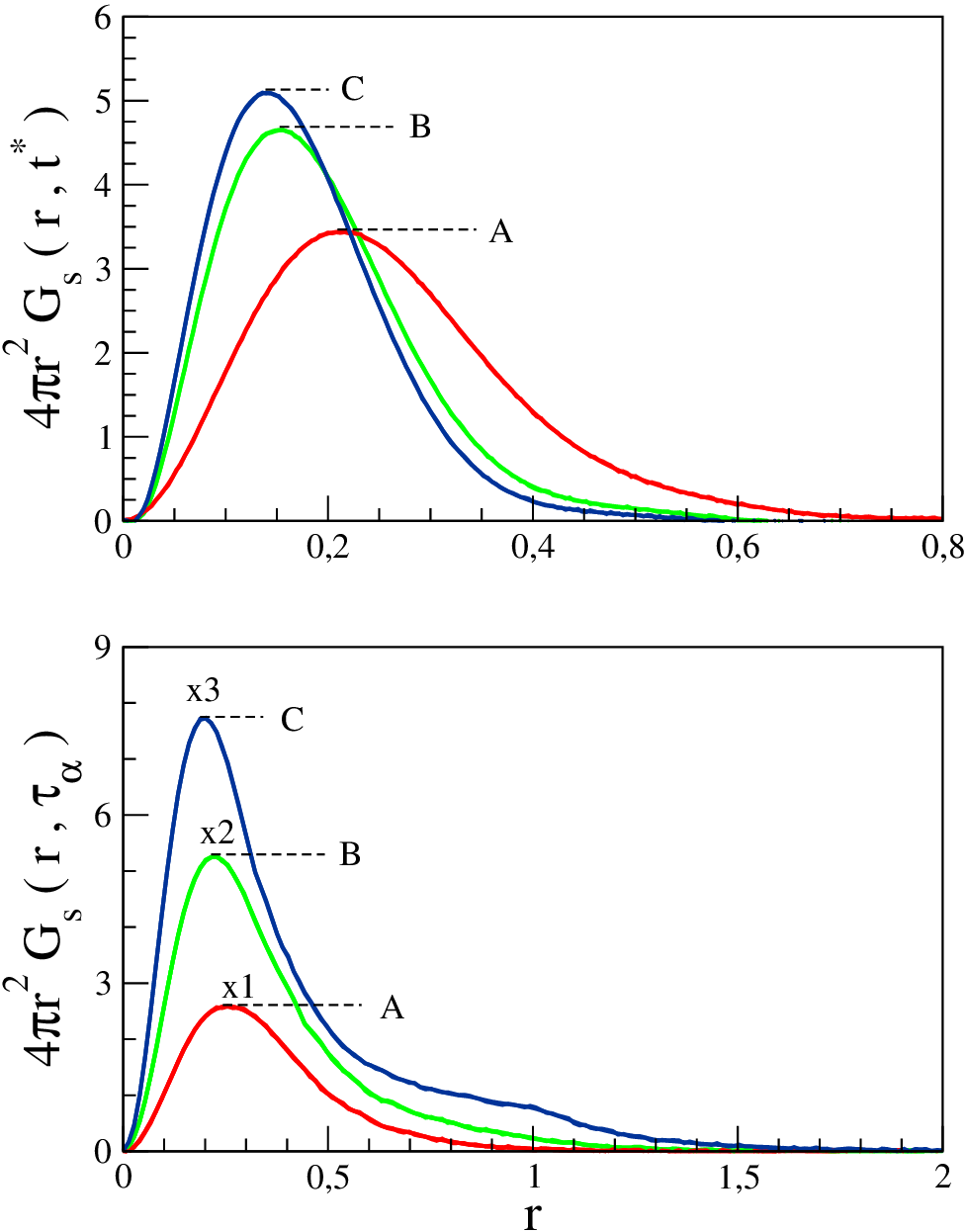}
\end{center}
\caption{Particle displacement distribution of the states of the sets A,B,C  
evaluated at 
 the rattling time $t = t^*$ (top) and the structural relaxation time
$t = \uptau_\alpha$ (bottom). Each curve results from the virtually {\it 
perfect}  superposition of the curves pertaining to the different states of each 
set. Note the characteristic bump at $r \sim 1$ due to the more apparent jump 
dynamics occurring for the states of the C set.
}
 \label{fig2}
\end{figure}

\section{Results and discussion}
\label{resultsdiscussion}
\subsection{Correlation between cage rattling and structural relaxation}
\label{dynamic}

A central quantity of interest is the distribution of the monomer displacements  
which is accounted for by  the self-part of the van
Hove function $G_s(\mathbf{r},t)$ \cite{Egelstaff:1992fk,HansenMcDonaldIIIEd}:

\begin{equation} \label{vh}
 G_s(\mathbf{r},t)=\frac{1}{N}\left\langle\sum_{i=1}^N \delta \left[ \mathbf{r}
+\mathbf{r}_i(0)- \mathbf{r}_i(t)\right] \right\rangle
\end{equation}

In isotropic liquids, the van Hove function depends on the modulus
r of $\mathbf{r}$. 
The second moment of $G_s(r,t)$ is related to the mean square displacement (MSD):

\begin{equation} \label{msd}
\left\langle r^2 (t) \right\rangle = 4\pi\int_0^\infty r^2 G_s(r,t) r^2 dr
\end{equation}

In order to characterize the cage fast dynamics we consider the MSD evaluated  
at the characteristic time scale $t^{\star}$ which is defined by the condition 
that the derivative $\Delta(t) \equiv \partial \log \langle r^{2}(t)\rangle/ 
\partial \log t $ is minimum at $t^{\star}$, i.e. $t^{\star}$ is the time when 
MSD changes the concavity in the log-log plot  \cite{OurNatPhys,lepoJCP09}. 
$t^{\star}$ is a measure of the trapping time of the particle  and corresponds 
to a few picoseconds in actual units. In the present model $t^{\star} \simeq 1 $ 
in MD units, irrespective of the physical state  \cite{OurNatPhys}, and ST-MSD 
is defined as $\langle u^2 \rangle = \langle r^2(t=t^*) \rangle $.

\begin{figure}[t]
\begin{center}
\includegraphics[width=0.5\linewidth]{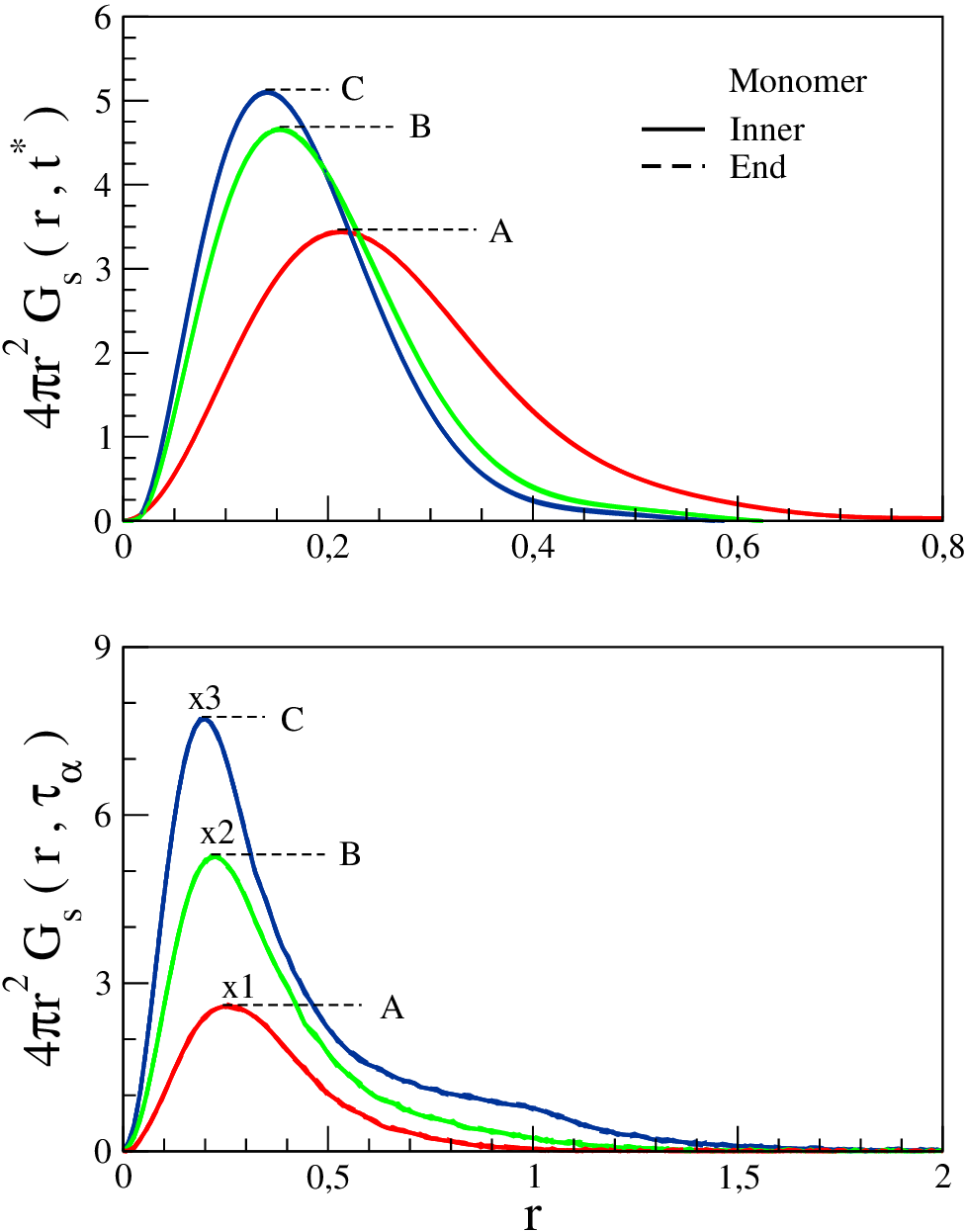}
\end{center}
\caption{Particle displacement distribution  of the inner and outer monomers  of 
the trimers for the sets of states A,B,C at  the rattling time $t = t^*$ (top) 
and  the structural relaxation time  $t = \uptau_\alpha$ (bottom). For each set 
of states all the curves are coinciding, namely,  the displacement distribution 
depends on neither the monomer position along the chain nor, for a given 
relaxation time, the state itself.}
 \label{fig2bis}
\end{figure}

\begin{figure}[t]
\begin{center}
\includegraphics[width=0.5\linewidth]{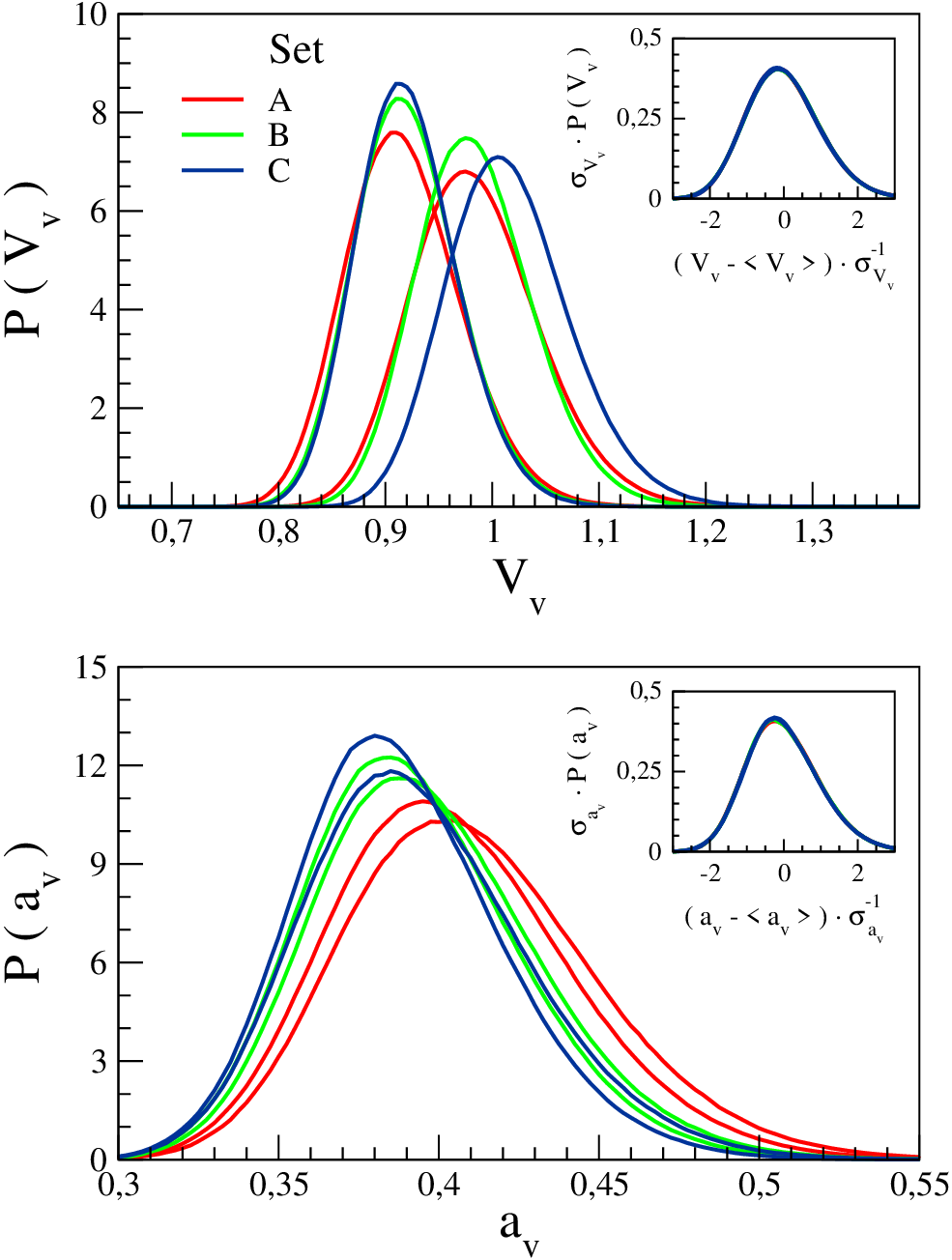} \\
\end{center}
\caption{Distribution of Voronoi polyhedra's volume $V_v$ (top) and
asphericity $a_v$(bottom), for set A (red curves), B (green curves) and C (blue
curves). The insets show that a single master curve is obtained by  
shifting the distributions by the average and scaled by the root-mean-square 
deviation, as noted before \cite{StarrEtAl02}.
}
 \label{fig3}
\end{figure}

The spatial Fourier transform of the self-part of the van Hove
function yields the self-part of the intermediate scattering function (ISF)
$F_s(q,t)$ \cite{HansenMcDonaldIIIEd}:

\begin{equation}  \label{isf}
F_s(q,t) = \int G_s(r,t) e^{-iq\cdot r} dr
\end{equation}

which, in an isotropic liquid, depends only on the modulus of the
wavevector $q = |\mathbf{q}| $. ISF is a useful tool to investigate the  
rearrangements of the spatial
structure over the length scale $~2\pi/q$. Since we are interested in the  
structural relaxation of the cage surrounding the tagged particle,
ISF is evaluated at $q = q_{max}$, where the maximum of the static
structure factor is located. Accordingly, the structural relaxation
time $\uptau_\alpha$ is defined by the equation $  
F_s(q_{max},\uptau_\alpha)=e^{-1} $.

Fig.\ref{fig1} shows ISF (top ) and MSD (middle) of the sets A,B,C.  In 
agreement with Eqs.\ref{vanhovescaling}, \ref{msd}, \ref{isf}, it is seen that 
states with equal ST-MSD $\langle u^2 \rangle$ have equal  $\tau_\alpha$ 
\cite{Puosi11,SpecialIssueJCP13,OurNatPhys}. Even more, states of the same set 
have coinciding MSD and ISF from times about  $t^*$ onward. This points to a 
strong correlation between the picosecond rattling motion in the cage and the 
structural relaxation (the deviations in the ballistic regime at very short 
times are due to the different temperatures). Careful studies show that the 
correlation is also present in the diffusive regime, i.e. states with equal 
ST-MSD have equal diffusivity \cite{Puosi11,SpecialIssueJCP13}.

The coincidence of MSD and ISF of states with equal $\tau_\alpha$ for  $t 
\gtrsim t^*$  reflects the scaling of the self-part of the van Hove function, 
Eq.\ref{vanhovescaling}. This is shown in Fig
\ref{fig2} which evidences  the coincidence of $G_s(r,t)$ with $t=t^*$ (top)  
and $t=\uptau_\alpha$
(bottom) for  states belonging to the same set. Note that the coincidence of   
$G_s(r,\tau_\alpha)$ of the states of the sets C extends up to $r \sim 1$ (the 
monomer diameter) where the characteristic bump due to the jump dynamics is 
apparent .

\begin{figure}[t]
\begin{center}
\includegraphics[width= 0.5 \linewidth]{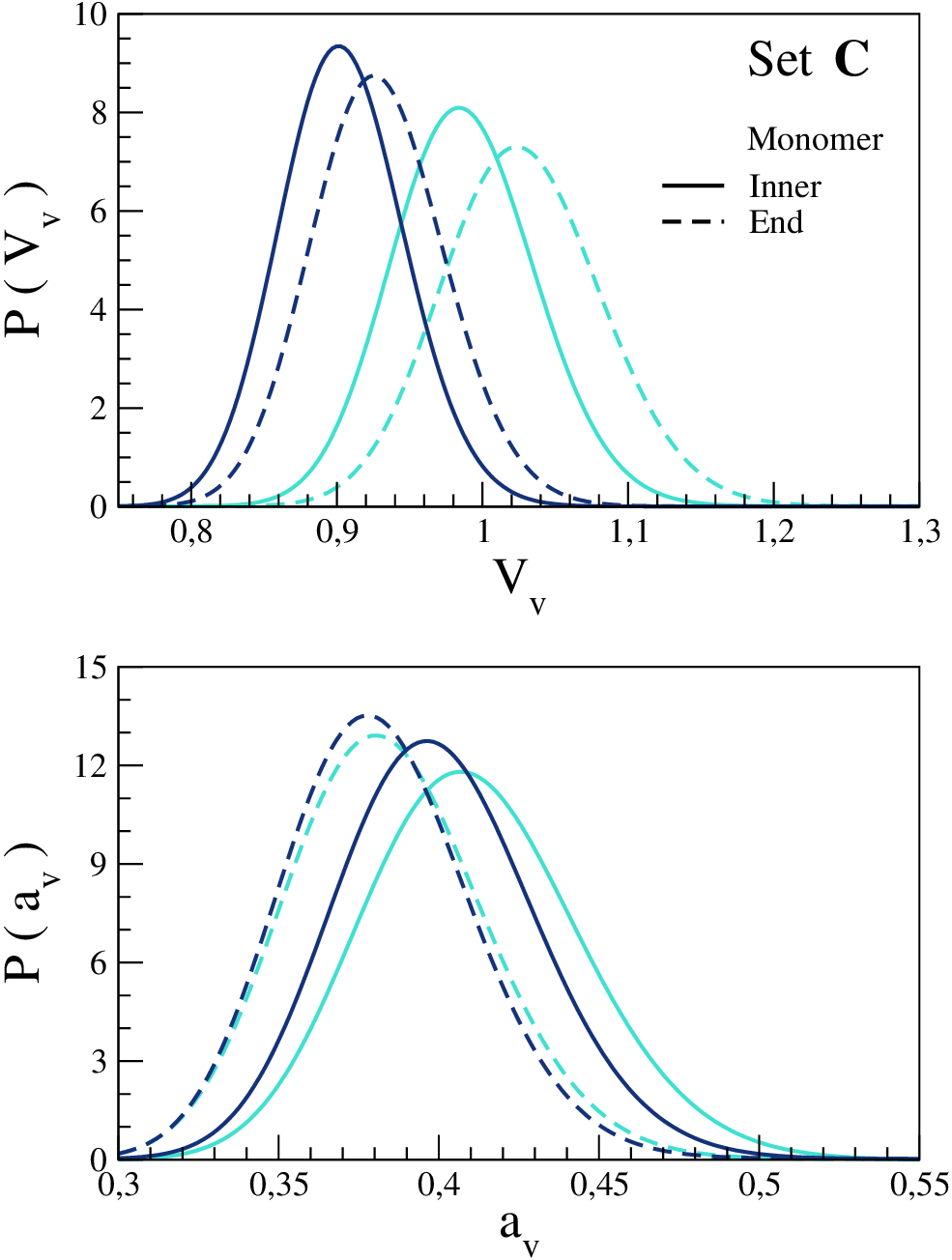}
\end{center}
\caption{Distribution of the Voronoi polyhedra's volume $V_v$ (top) and  
asphericity $a_v$(bottom) for the inner and the end monomers of the trimers of 
the states of the set C. Thick blue: (1.086,6,12,0.63), light blue : 
(0.984,6,12,0.33).}
 \label{fig5}
\end{figure}

The displacement distribution in Fig.\ref{fig2} is averaged over {\it all}  the 
monomers.  We now prove that, actually, the shape of the distribution is 
independent of  the specific monomer. To this aim, Fig.\ref{fig2bis} shows  the 
van Hove functions of the inner and the outer monomers of the trimers for the 
sets of states A,B,C. It is seen that they are virtually coincident  at both 
short ($t = t^*$) and long ($t = \uptau_\alpha$) times and {\it independent} of 
the physical state for a given relaxation time. 

\subsection{Local geometry: Voronoi polyhedra}
\label{voronoi}

\begin{figure}[t]
\begin{center}
\includegraphics[width= 0.5 \linewidth]{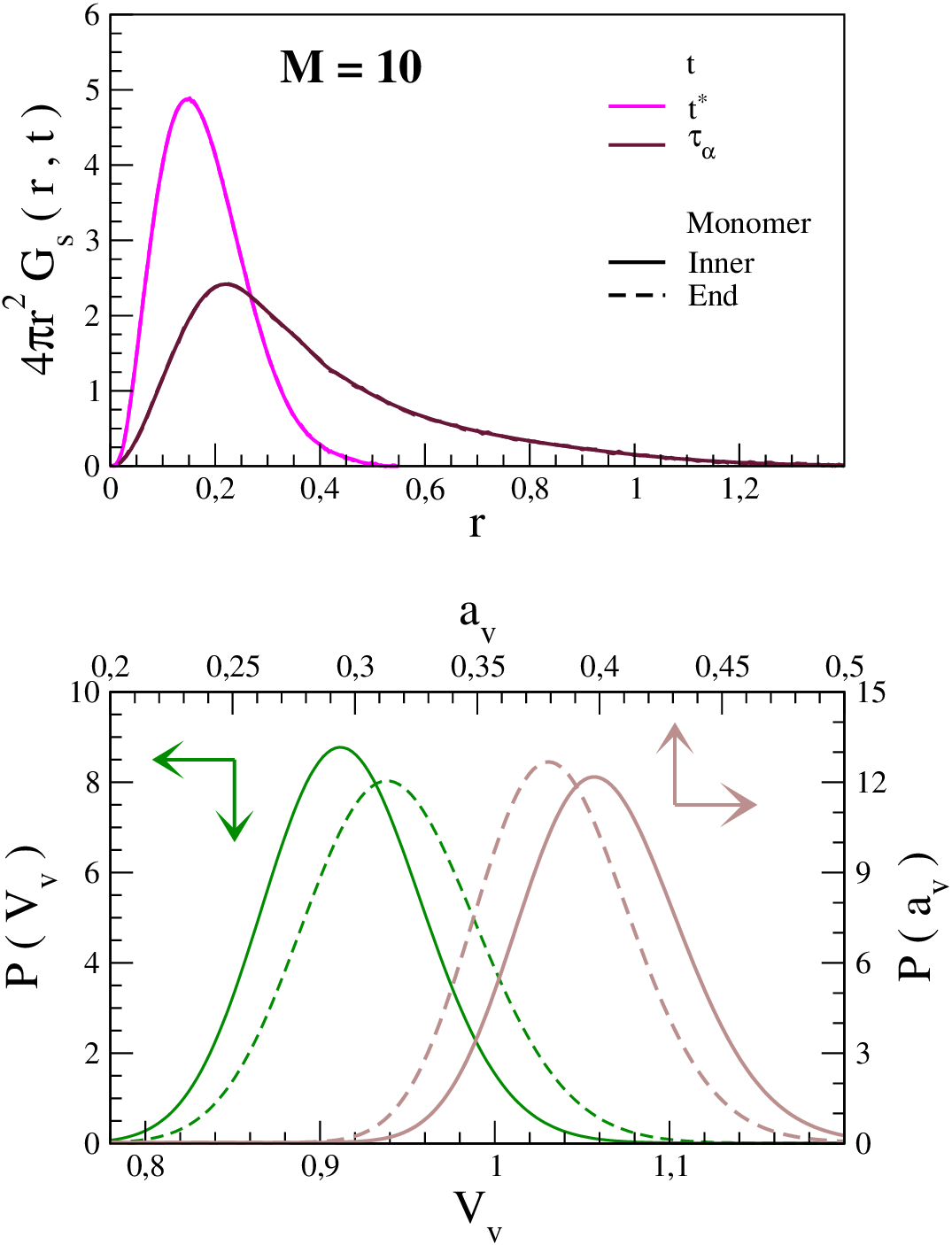}
\end{center}
\caption{Dynamics and local geometry of the two end and the two innermost  
monomers of a single chain molecule for the melt of decamers. Top: distribution 
of the monomer displacements at short and long times. Notice that the 
displacement distributions of the end and the inner monomers are coinciding. 
Bottom: distributions of the VPs volume  (green) and asphericity  (brown).  }
 \label{fig6}
\end{figure}

To characterize the local geometry of the neighborhood of a tagged particle  we 
perform a space tessellation in terms of the Voronoi polyhedra. Given a 
particular arrangement of the monomers,  the Voronoi polyhedron (VP)  
surrounding a tagged monomer encloses all the points which are closer to it 
than to any other one.  In particular, we are interested in both their  volume 
$V_v$  and  their asphericity $a_v$, Eq.\ref{aspher}.

Fig.\ref{fig3} shows the distributions of the volume (top) and the asphericity  
(bottom) of the VPs for the three sets of states under investigation. It is 
apparent that states of the same set do not have the same distribution of local 
geometries. Remind that states belonging to the same set have coinciding 
monomer displacement distribution  at both short and long times 
(Fig.\ref{fig2}). The insets of Fig.\ref{fig3} show that the distributions 
of both the  volume and the asphericity of the VPs  collapse to a single master 
curve by following Starr et al  \cite{StarrEtAl02}. This suggests the existence 
of a single underlying distribution.

To provide more insight, we considered the C set of states and analyzed in a  
separate way the volume and the asphericity distributions of the end and the 
inner monomers. The results are plotted  in Fig.\ref{fig5}. It is seen (top 
panel)  that end monomers have extra Voronoi volume with respect to the 
inner monomers, as it is well known since long times \cite{Bueche} and observed 
in previous MD work \cite{BarbieriEtAl2004}. Furthermore, VPs of the end 
monomers are more spherical. In spite of the different local geometries around 
the end and the inner monomers their displacement distributions are identical 
at both short and long times (Fig.\ref{fig2bis}). This results is in agreement 
with MD studies on a 2D glass-forming alloy showing that having a larger  
Voronoi volume does not cause a particle to exhibit larger amplitude 
fluctuations in position \cite{HarrowellFreeVolume06}.

It must be pointed out that the above results are not limited to trimers -  
where some correlation between the motion of the end and the inner monomers 
arises from  their direct bonding - but it is also observed in longer 
molecules. In fact, Fig.\ref{fig6} shows that also in a melt of decamers the 
displacement distribution of the two end and the two innermost monomers {\it 
coincide}, whereas the local geometry around them is {\it different}.

Our findings suggest that the fast motion of a particle within the cage of  the 
neighbors is a non-local process. We propose that it results from the 
collective motion of particles in a region extending more than the first 
neighbours region. In fact, equal-time, i.e. simultaneous, particle motion 
correlating over the next-nearest neighbours even at the short (picosecond) 
time scales has been  evidenced 
\cite{PuosiLepoJCPCor12,PuosiLepoJCPCor12_Erratum}. This is consistent 
with the evidence of quasi-local soft modes, i.e. with an extended component, 
in MD simulations of  2D binary mixtures and the finding that the regions of 
motion of the quasi-local soft modes exhibit striking correlation with the 
regions of high Debye-Waller factor  
\cite{Harrowell06,Harrowell_NP08,HarrowellJCP09}.

The conclusion that the influence of structure on dynamics is  weak on short  
length scale has been reached before by Berthier and Jack, who predicted that 
it  becomes much stronger on long length scale \cite{BerthierJackPRE07}. From 
this respect, the evidence of strong correlations between the (long wavelength) 
elasticity and the fast and slow dynamics deserves further consideration  
\cite{Puosi12,Elastico2}.

\section{Conclusions}
\label{conclusions}
A supercooled liquid of linear molecules has been investigated by MD 
simulations.  The emphasis is on the possible correlation between the local 
order and the dynamics at short and long times, as probed by cage rattling and 
structural relaxation, respectively. Two measures of the local order, i.e. the 
VP volume and asphericity have no clear correlation with both fast and slow 
dynamics.  In particular, it is found that the excess Voronoi volume 
surrounding the end monomers does not enhance their dynamics neither at short 
nor at long times. It is concluded that  fast dynamics is a non-local process 
extending farther than the first neighbor shell.


\begin{thebibliography}{89}
\expandafter\ifx\csname natexlab\endcsname\relax\def\natexlab#1{#1}\fi
\expandafter\ifx\csname bibnamefont\endcsname\relax
  \def\bibnamefont#1{#1}\fi
\expandafter\ifx\csname bibfnamefont\endcsname\relax
  \def\bibfnamefont#1{#1}\fi
\expandafter\ifx\csname citenamefont\endcsname\relax
  \def\citenamefont#1{#1}\fi
\expandafter\ifx\csname url\endcsname\relax
  \def\url#1{\texttt{#1}}\fi
\expandafter\ifx\csname urlprefix\endcsname\relax\def\urlprefix{URL }\fi
\providecommand{\bibinfo}[2]{#2}
\providecommand{\eprint}[2][]{\url{#2}}

\bibitem[{\citenamefont{ten Wolde et~al.}(2005)\citenamefont{ten Wolde,
  Ruiz-Montero, and Frenkel}}]{FrenkelPRL95}
\bibinfo{author}{\bibfnamefont{P.~R.} \bibnamefont{ten Wolde}},
  \bibinfo{author}{\bibfnamefont{M.~J.} \bibnamefont{Ruiz-Montero}},
  \bibnamefont{and} \bibinfo{author}{\bibfnamefont{D.}~\bibnamefont{Frenkel}},
  \bibinfo{journal}{Phys. Rev. Lett.} \textbf{\bibinfo{volume}{75}},
  \bibinfo{pages}{2714} (\bibinfo{year}{2005}).

\bibitem[{\citenamefont{Kansal et~al.}(2002)\citenamefont{Kansal, Torquato, and
  Stillinger}}]{TorquatoStilliPRE02}
\bibinfo{author}{\bibfnamefont{A.~R.} \bibnamefont{Kansal}},
  \bibinfo{author}{\bibfnamefont{S.}~\bibnamefont{Torquato}}, \bibnamefont{and}
  \bibinfo{author}{\bibfnamefont{F.~H.} \bibnamefont{Stillinger}},
  \bibinfo{journal}{Phys. Rev. E} \textbf{\bibinfo{volume}{66}},
  \bibinfo{pages}{041109} (\bibinfo{year}{2002}).

\bibitem[{\citenamefont{Aste}(2005)}]{AsteJPCM05}
\bibinfo{author}{\bibfnamefont{T.}~\bibnamefont{Aste}}, \bibinfo{journal}{J.
  Phys.: Condens. Matter} \textbf{\bibinfo{volume}{17}}, \bibinfo{pages}{S2361}
  (\bibinfo{year}{2005}).

\bibitem[{\citenamefont{Bernini et~al.}(2013)\citenamefont{Bernini, Puosi,
  Barucco, and Leporini}}]{LocalOrderJCP13}
\bibinfo{author}{\bibfnamefont{S.}~\bibnamefont{Bernini}},
  \bibinfo{author}{\bibfnamefont{F.}~\bibnamefont{Puosi}},
  \bibinfo{author}{\bibfnamefont{M.}~\bibnamefont{Barucco}}, \bibnamefont{and}
  \bibinfo{author}{\bibfnamefont{D.}~\bibnamefont{Leporini}},
  \bibinfo{journal}{J. Chem. Phys.} \textbf{\bibinfo{volume}{139}},
  \bibinfo{pages}{184501} (\bibinfo{year}{2013}).

\bibitem[{\citenamefont{Chandler and Garrahan}(2010)}]{ChandlerGarra10}
\bibinfo{author}{\bibfnamefont{D.}~\bibnamefont{Chandler}} \bibnamefont{and}
  \bibinfo{author}{\bibfnamefont{J.~P.} \bibnamefont{Garrahan}},
  \bibinfo{journal}{Annu. Rev. Phys. Chem.} \textbf{\bibinfo{volume}{61}},
  \bibinfo{pages}{191} (\bibinfo{year}{2010}).

\bibitem[{\citenamefont{Berthier and Biroli}(2011)}]{BerthierBiroliRMP11}
\bibinfo{author}{\bibfnamefont{L.}~\bibnamefont{Berthier}} \bibnamefont{and}
  \bibinfo{author}{\bibfnamefont{G.}~\bibnamefont{Biroli}},
  \bibinfo{journal}{Rev. Mod. Phys.} \textbf{\bibinfo{volume}{83}},
  \bibinfo{pages}{587} (\bibinfo{year}{2011}).

\bibitem[{\citenamefont{Adam and Gibbs}(1965)}]{AdamGibbs65}
\bibinfo{author}{\bibfnamefont{G.}~\bibnamefont{Adam}} \bibnamefont{and}
  \bibinfo{author}{\bibfnamefont{J.~H.} \bibnamefont{Gibbs}},
  \bibinfo{journal}{J. Chem. Phys.} \textbf{\bibinfo{volume}{43}},
  \bibinfo{pages}{139} (\bibinfo{year}{1965}).

\bibitem[{\citenamefont{Dudowicz et~al.}(2008)\citenamefont{Dudowicz, Freed,
  and Douglas}}]{DudowiczEtAl08}
\bibinfo{author}{\bibfnamefont{J.}~\bibnamefont{Dudowicz}},
  \bibinfo{author}{\bibfnamefont{K.~F.} \bibnamefont{Freed}}, \bibnamefont{and}
  \bibinfo{author}{\bibfnamefont{J.~F.} \bibnamefont{Douglas}},
  \bibinfo{journal}{Adv. Chem. Phys.} \textbf{\bibinfo{volume}{137}},
  \bibinfo{pages}{125} (\bibinfo{year}{2008}).

\bibitem[{\citenamefont{Gibbs and DiMarzio}(1958)}]{GibbsDiMarzio58}
\bibinfo{author}{\bibfnamefont{J.~H.} \bibnamefont{Gibbs}} \bibnamefont{and}
  \bibinfo{author}{\bibfnamefont{E.~A.} \bibnamefont{DiMarzio}},
  \bibinfo{journal}{J. Chem. Phys.} \textbf{\bibinfo{volume}{28}},
  \bibinfo{pages}{373} (\bibinfo{year}{1958}).

\bibitem[{\citenamefont{G\"otze}(2008)}]{GotzeBook}
\bibinfo{author}{\bibfnamefont{W.}~\bibnamefont{G\"otze}},
  \emph{\bibinfo{title}{Complex Dynamics of Glass-Forming Liquids: A
  Mode-Coupling Theory}} (\bibinfo{publisher}{Oxford University Press, Oxford},
  \bibinfo{year}{2008}).

\bibitem[{\citenamefont{Chen et~al.}(2010)\citenamefont{Chen, Saltzman, and
  Schweizer}}]{SchweizerAnnRev10}
\bibinfo{author}{\bibfnamefont{K.}~\bibnamefont{Chen}},
  \bibinfo{author}{\bibfnamefont{E.~J.} \bibnamefont{Saltzman}},
  \bibnamefont{and} \bibinfo{author}{\bibfnamefont{K.~S.}
  \bibnamefont{Schweizer}}, \bibinfo{journal}{Annu. Rev. Condens. Matter Phys.}
  \textbf{\bibinfo{volume}{1}}, \bibinfo{pages}{277} (\bibinfo{year}{2010}).

\bibitem[{\citenamefont{Lubchenko and Wolynes}(2007)}]{WolynesRFOT07}
\bibinfo{author}{\bibfnamefont{V.}~\bibnamefont{Lubchenko}} \bibnamefont{and}
  \bibinfo{author}{\bibfnamefont{P.~G.} \bibnamefont{Wolynes}},
  \bibinfo{journal}{Annu. Rev. Phys. Chem.} \textbf{\bibinfo{volume}{58}},
  \bibinfo{pages}{235} (\bibinfo{year}{2007}).

\bibitem[{\citenamefont{Tarjus et~al.}(2005)\citenamefont{Tarjus, Kivelson,
  Nussinov, and Viot}}]{TarjusJPCM05}
\bibinfo{author}{\bibfnamefont{G.}~\bibnamefont{Tarjus}},
  \bibinfo{author}{\bibfnamefont{S.~A.} \bibnamefont{Kivelson}},
  \bibinfo{author}{\bibfnamefont{Z.}~\bibnamefont{Nussinov}}, \bibnamefont{and}
  \bibinfo{author}{\bibfnamefont{P.}~\bibnamefont{Viot}}, \bibinfo{journal}{J.
  Phys.: Condens. Matter} \textbf{\bibinfo{volume}{17}}, \bibinfo{pages}{R1143}
  (\bibinfo{year}{2005}).

\bibitem[{\citenamefont{Dyre}(2006)}]{Dyre06}
\bibinfo{author}{\bibfnamefont{J.~C.} \bibnamefont{Dyre}},
  \bibinfo{journal}{Rev. Mod. Phys.} \textbf{\bibinfo{volume}{78}},
  \bibinfo{pages}{953} (\bibinfo{year}{2006}).

\bibitem[{\citenamefont{Puosi and Leporini}(2012{\natexlab{a}})}]{Puosi12}
\bibinfo{author}{\bibfnamefont{F.}~\bibnamefont{Puosi}} \bibnamefont{and}
  \bibinfo{author}{\bibfnamefont{D.}~\bibnamefont{Leporini}},
  \bibinfo{journal}{J. Chem. Phys.} \textbf{\bibinfo{volume}{136}},
  \bibinfo{pages}{041104} (\bibinfo{year}{2012}{\natexlab{a}}).

\bibitem[{\citenamefont{Capponi et~al.}(2012)\citenamefont{Capponi, Napolitano,
  and W\"ubbenhorst}}]{NapolitanoNatCom12}
\bibinfo{author}{\bibfnamefont{S.}~\bibnamefont{Capponi}},
  \bibinfo{author}{\bibfnamefont{S.}~\bibnamefont{Napolitano}},
  \bibnamefont{and}
  \bibinfo{author}{\bibfnamefont{M.}~\bibnamefont{W\"ubbenhorst}},
  \bibinfo{journal}{Nat. Commun.} \textbf{\bibinfo{volume}{3}},
  \bibinfo{pages}{1233} (\bibinfo{year}{2012}).

\bibitem[{\citenamefont{Singh et~al.}(2013)\citenamefont{Singh, Ediger, and
  de~Pablo}}]{EdigerDePabloNatMat13}
\bibinfo{author}{\bibfnamefont{S.}~\bibnamefont{Singh}},
  \bibinfo{author}{\bibfnamefont{M.~D.} \bibnamefont{Ediger}},
  \bibnamefont{and} \bibinfo{author}{\bibfnamefont{J.~J.}
  \bibnamefont{de~Pablo}}, \bibinfo{journal}{Nat. Mater.}
  \textbf{\bibinfo{volume}{12}}, \bibinfo{pages}{139} (\bibinfo{year}{2013}).

\bibitem[{\citenamefont{Speck et~al.}(2012)\citenamefont{Speck, Malins, and
  Royall}}]{RoyallPRL12}
\bibinfo{author}{\bibfnamefont{T.}~\bibnamefont{Speck}},
  \bibinfo{author}{\bibfnamefont{A.}~\bibnamefont{Malins}}, \bibnamefont{and}
  \bibinfo{author}{\bibfnamefont{C.~P.} \bibnamefont{Royall}},
  \bibinfo{journal}{Phys. Rev. Lett.} \textbf{\bibinfo{volume}{109}},
  \bibinfo{pages}{195703} (\bibinfo{year}{2012}).

\bibitem[{\citenamefont{Barbieri
  et~al.}(2004{\natexlab{a}})\citenamefont{Barbieri, Gorini, and
  Leporini}}]{BarbieriGoriniPRE04}
\bibinfo{author}{\bibfnamefont{A.}~\bibnamefont{Barbieri}},
  \bibinfo{author}{\bibfnamefont{G.}~\bibnamefont{Gorini}}, \bibnamefont{and}
  \bibinfo{author}{\bibfnamefont{D.}~\bibnamefont{Leporini}},
  \bibinfo{journal}{Phys. Rev. E} \textbf{\bibinfo{volume}{69}},
  \bibinfo{pages}{061509} (\bibinfo{year}{2004}{\natexlab{a}}).

\bibitem[{\citenamefont{Conrad et~al.}(2005)\citenamefont{Conrad, Starr, and
  Weitz}}]{StarrWeitz05}
\bibinfo{author}{\bibfnamefont{J.~C.} \bibnamefont{Conrad}},
  \bibinfo{author}{\bibfnamefont{F.~W.} \bibnamefont{Starr}}, \bibnamefont{and}
  \bibinfo{author}{\bibfnamefont{D.~A.} \bibnamefont{Weitz}},
  \bibinfo{journal}{J.Phys. Chem. B} \textbf{\bibinfo{volume}{109}},
  \bibinfo{pages}{21235} (\bibinfo{year}{2005}).

\bibitem[{\citenamefont{Royall et~al.}(2008)\citenamefont{Royall, Williams,
  Ohtsuka, and Tanaka}}]{TanakaNatMater08}
\bibinfo{author}{\bibfnamefont{C.~P.} \bibnamefont{Royall}},
  \bibinfo{author}{\bibfnamefont{S.~R.} \bibnamefont{Williams}},
  \bibinfo{author}{\bibfnamefont{T.}~\bibnamefont{Ohtsuka}}, \bibnamefont{and}
  \bibinfo{author}{\bibfnamefont{H.}~\bibnamefont{Tanaka}},
  \bibinfo{journal}{Nat. Mater.} \textbf{\bibinfo{volume}{7}},
  \bibinfo{pages}{556} (\bibinfo{year}{2008}).

\bibitem[{\citenamefont{Leocmach and Tanaka}(2012)}]{TanakaNatCom12}
\bibinfo{author}{\bibfnamefont{M.}~\bibnamefont{Leocmach}} \bibnamefont{and}
  \bibinfo{author}{\bibfnamefont{H.}~\bibnamefont{Tanaka}},
  \bibinfo{journal}{Nat. Commun.} \textbf{\bibinfo{volume}{3}},
  \bibinfo{pages}{974} (\bibinfo{year}{2012}).

\bibitem[{\citenamefont{Jain and de~Pablo}(2005)}]{DePabloJCP05}
\bibinfo{author}{\bibfnamefont{T.~S.} \bibnamefont{Jain}} \bibnamefont{and}
  \bibinfo{author}{\bibfnamefont{J.}~\bibnamefont{de~Pablo}},
  \bibinfo{journal}{J.Chem.Phys.} \textbf{\bibinfo{volume}{122}},
  \bibinfo{pages}{174515} (\bibinfo{year}{2005}).

\bibitem[{\citenamefont{Iacovella et~al.}(2007)\citenamefont{Iacovella, Keys,
  Horsch, and Glotzer}}]{GlotzerPRE07}
\bibinfo{author}{\bibfnamefont{C.~R.} \bibnamefont{Iacovella}},
  \bibinfo{author}{\bibfnamefont{A.~S.} \bibnamefont{Keys}},
  \bibinfo{author}{\bibfnamefont{M.~A.} \bibnamefont{Horsch}},
  \bibnamefont{and} \bibinfo{author}{\bibfnamefont{S.~C.}
  \bibnamefont{Glotzer}}, \bibinfo{journal}{Phys. Rev. E}
  \textbf{\bibinfo{volume}{75}}, \bibinfo{pages}{040801(R)}
  (\bibinfo{year}{2007}).

\bibitem[{\citenamefont{Karayiannis et~al.}(2009)\citenamefont{Karayiannis,
  Foteinopoulou, and Laso}}]{LasoJCP09}
\bibinfo{author}{\bibfnamefont{N.~C.} \bibnamefont{Karayiannis}},
  \bibinfo{author}{\bibfnamefont{K.}~\bibnamefont{Foteinopoulou}},
  \bibnamefont{and} \bibinfo{author}{\bibfnamefont{M.}~\bibnamefont{Laso}},
  \bibinfo{journal}{J. Chem. Phys.} \textbf{\bibinfo{volume}{130}},
  \bibinfo{pages}{164908} (\bibinfo{year}{2009}).

\bibitem[{\citenamefont{Schnell et~al.}(2011)\citenamefont{Schnell, Meyer,
  Fond, Wittmer, and Baschnagel}}]{BaschnagelEPJE11}
\bibinfo{author}{\bibfnamefont{B.}~\bibnamefont{Schnell}},
  \bibinfo{author}{\bibfnamefont{H.}~\bibnamefont{Meyer}},
  \bibinfo{author}{\bibfnamefont{C.}~\bibnamefont{Fond}},
  \bibinfo{author}{\bibfnamefont{J.}~\bibnamefont{Wittmer}}, \bibnamefont{and}
  \bibinfo{author}{\bibfnamefont{J.}~\bibnamefont{Baschnagel}},
  \bibinfo{journal}{Eur. Phys. J. E} \textbf{\bibinfo{volume}{34}},
  \bibinfo{pages}{97} (\bibinfo{year}{2011}).

\bibitem[{\citenamefont{Asai et~al.}(2011)\citenamefont{Asai, Shibayama, and
  Koike}}]{MakotoMM11}
\bibinfo{author}{\bibfnamefont{M.}~\bibnamefont{Asai}},
  \bibinfo{author}{\bibfnamefont{M.}~\bibnamefont{Shibayama}},
  \bibnamefont{and} \bibinfo{author}{\bibfnamefont{Y.}~\bibnamefont{Koike}},
  \bibinfo{journal}{Macromolecules} \textbf{\bibinfo{volume}{44}},
  \bibinfo{pages}{6615} (\bibinfo{year}{2011}).

\bibitem[{\citenamefont{Larini et~al.}(2005)\citenamefont{Larini, Barbieri,
  Prevosto, Rolla, and Leporini}}]{LariniCrystJPCM05}
\bibinfo{author}{\bibfnamefont{L.}~\bibnamefont{Larini}},
  \bibinfo{author}{\bibfnamefont{A.}~\bibnamefont{Barbieri}},
  \bibinfo{author}{\bibfnamefont{D.}~\bibnamefont{Prevosto}},
  \bibinfo{author}{\bibfnamefont{P.~A.} \bibnamefont{Rolla}}, \bibnamefont{and}
  \bibinfo{author}{\bibfnamefont{D.}~\bibnamefont{Leporini}},
  \bibinfo{journal}{J. Phys.: Condens. Matter} \textbf{\bibinfo{volume}{17}},
  \bibinfo{pages}{L199} (\bibinfo{year}{2005}).

\bibitem[{\citenamefont{Ediger and Harrowell}(2012)}]{EdigerHarrowell12}
\bibinfo{author}{\bibfnamefont{M.~D.} \bibnamefont{Ediger}} \bibnamefont{and}
  \bibinfo{author}{\bibfnamefont{P.}~\bibnamefont{Harrowell}},
  \bibinfo{journal}{J. Chem. Phys.} \textbf{\bibinfo{volume}{137}},
  \bibinfo{pages}{080901} (\bibinfo{year}{2012}).

\bibitem[{\citenamefont{Angell et~al.}(2000)\citenamefont{Angell, Ngai,
  McKenna, McMillan, and S.W.Martin}}]{AngelNgai00}
\bibinfo{author}{\bibfnamefont{C.~A.} \bibnamefont{Angell}},
  \bibinfo{author}{\bibfnamefont{K.~L.} \bibnamefont{Ngai}},
  \bibinfo{author}{\bibfnamefont{G.~B.} \bibnamefont{McKenna}},
  \bibinfo{author}{\bibfnamefont{P.}~\bibnamefont{McMillan}}, \bibnamefont{and}
  \bibinfo{author}{\bibnamefont{S.W.Martin}}, \bibinfo{journal}{J. Appl. Phys.}
  \textbf{\bibinfo{volume}{88}}, \bibinfo{pages}{3113} (\bibinfo{year}{2000}).

\bibitem[{\citenamefont{Debenedetti and Stillinger}(2001)}]{DebeStilli2001}
\bibinfo{author}{\bibfnamefont{P.~G.} \bibnamefont{Debenedetti}}
  \bibnamefont{and} \bibinfo{author}{\bibfnamefont{F.~H.}
  \bibnamefont{Stillinger}}, \bibinfo{journal}{Nature}
  \textbf{\bibinfo{volume}{410}}, \bibinfo{pages}{259} (\bibinfo{year}{2001}).

\bibitem[{\citenamefont{Richert}(2002)}]{Richert02}
\bibinfo{author}{\bibfnamefont{R.}~\bibnamefont{Richert}}, \bibinfo{journal}{J.
  Phys.: Condens. Matter} \textbf{\bibinfo{volume}{14}}, \bibinfo{pages}{R703}
  (\bibinfo{year}{2002}).

\bibitem[{\citenamefont{Monthus and Bouchaud}(1996)}]{MonthBouch96}
\bibinfo{author}{\bibfnamefont{C.}~\bibnamefont{Monthus}} \bibnamefont{and}
  \bibinfo{author}{\bibfnamefont{J.-P.} \bibnamefont{Bouchaud}},
  \bibinfo{journal}{J. Phys. A: Math. Gen.} \textbf{\bibinfo{volume}{29}},
  \bibinfo{pages}{3847} (\bibinfo{year}{1996}).

\bibitem[{\citenamefont{Tobolsky et~al.}(1943)\citenamefont{Tobolsky, Powell,
  and Eyring}}]{TobolskyEtAl43}
\bibinfo{author}{\bibfnamefont{A.}~\bibnamefont{Tobolsky}},
  \bibinfo{author}{\bibfnamefont{R.~E.} \bibnamefont{Powell}},
  \bibnamefont{and} \bibinfo{author}{\bibfnamefont{H.}~\bibnamefont{Eyring}},
  in \emph{\bibinfo{booktitle}{Frontiers in Chemistry}}, edited by
  \bibinfo{editor}{\bibfnamefont{R.~E.} \bibnamefont{Burk}} \bibnamefont{and}
  \bibinfo{editor}{\bibfnamefont{O.}~\bibnamefont{Grummit}}
  (\bibinfo{publisher}{Interscience}, \bibinfo{address}{New York},
  \bibinfo{year}{1943}), vol.~\bibinfo{volume}{1}, pp.
  \bibinfo{pages}{125--190}.

\bibitem[{\citenamefont{Angell}(1995)}]{Angell95}
\bibinfo{author}{\bibfnamefont{C.~A.} \bibnamefont{Angell}},
  \bibinfo{journal}{Science} \textbf{\bibinfo{volume}{267}},
  \bibinfo{pages}{1924} (\bibinfo{year}{1995}).

\bibitem[{\citenamefont{Hall and Wolynes}(1987)}]{HallWoly87}
\bibinfo{author}{\bibfnamefont{R.~W.} \bibnamefont{Hall}} \bibnamefont{and}
  \bibinfo{author}{\bibfnamefont{P.~G.} \bibnamefont{Wolynes}},
  \bibinfo{journal}{J. Chem. Phys.} \textbf{\bibinfo{volume}{86}},
  \bibinfo{pages}{2943} (\bibinfo{year}{1987}).

\bibitem[{\citenamefont{Dyre et~al.}(1996)\citenamefont{Dyre, Olsen, and
  Christensen}}]{Dyre96}
\bibinfo{author}{\bibfnamefont{J.~C.} \bibnamefont{Dyre}},
  \bibinfo{author}{\bibfnamefont{N.~B.} \bibnamefont{Olsen}}, \bibnamefont{and}
  \bibinfo{author}{\bibfnamefont{T.}~\bibnamefont{Christensen}},
  \bibinfo{journal}{Phys. Rev. B} \textbf{\bibinfo{volume}{53}},
  \bibinfo{pages}{2171} (\bibinfo{year}{1996}).

\bibitem[{\citenamefont{Martinez and Angell}(2001)}]{MarAngell01}
\bibinfo{author}{\bibfnamefont{L.-M.} \bibnamefont{Martinez}} \bibnamefont{and}
  \bibinfo{author}{\bibfnamefont{C.~A.} \bibnamefont{Angell}},
  \bibinfo{journal}{Nature} \textbf{\bibinfo{volume}{410}},
  \bibinfo{pages}{663} (\bibinfo{year}{2001}).

\bibitem[{\citenamefont{Ngai}(2004)}]{Ngai04}
\bibinfo{author}{\bibfnamefont{K.~L.} \bibnamefont{Ngai}},
  \bibinfo{journal}{Phil. Mag.} \textbf{\bibinfo{volume}{84}},
  \bibinfo{pages}{1341} (\bibinfo{year}{2004}).

\bibitem[{\citenamefont{Ngai}(2000)}]{Ngai00}
\bibinfo{author}{\bibfnamefont{K.~L.} \bibnamefont{Ngai}}, \bibinfo{journal}{J.
  Non-Cryst. Solids} \textbf{\bibinfo{volume}{275}}, \bibinfo{pages}{7}
  (\bibinfo{year}{2000}).

\bibitem[{\citenamefont{Simmons et~al.}(2012)\citenamefont{Simmons, Cicerone,
  Zhong, Tyagic, and Douglas}}]{DouglasCiceroneSoftMatter12}
\bibinfo{author}{\bibfnamefont{D.~S.} \bibnamefont{Simmons}},
  \bibinfo{author}{\bibfnamefont{M.~T.} \bibnamefont{Cicerone}},
  \bibinfo{author}{\bibfnamefont{Q.}~\bibnamefont{Zhong}},
  \bibinfo{author}{\bibfnamefont{M.}~\bibnamefont{Tyagic}}, \bibnamefont{and}
  \bibinfo{author}{\bibfnamefont{J.~F.} \bibnamefont{Douglas}},
  \bibinfo{journal}{Soft Matter} \textbf{\bibinfo{volume}{8}},
  \bibinfo{pages}{11455} (\bibinfo{year}{2012}).

\bibitem[{\citenamefont{Angell}(1968)}]{Angell68}
\bibinfo{author}{\bibfnamefont{C.~A.} \bibnamefont{Angell}},
  \bibinfo{journal}{J. Am. Chem. Soc.} \textbf{\bibinfo{volume}{86}},
  \bibinfo{pages}{117} (\bibinfo{year}{1968}).

\bibitem[{\citenamefont{S.V.Nemilov}(1968)}]{Nemilov68}
\bibinfo{author}{\bibnamefont{S.V.Nemilov}}, \bibinfo{journal}{Russ. J. Phys.
  Chem.} \textbf{\bibinfo{volume}{42}}, \bibinfo{pages}{726}
  (\bibinfo{year}{1968}).

\bibitem[{\citenamefont{Shao and Angell}(1995)}]{Angell95B}
\bibinfo{author}{\bibfnamefont{J.}~\bibnamefont{Shao}} \bibnamefont{and}
  \bibinfo{author}{\bibfnamefont{C.~A.} \bibnamefont{Angell}}, in
  \emph{\bibinfo{booktitle}{Proc. XVIIth International Congress on Glass,
  Beijing}} (\bibinfo{organization}{Chinese Ceramic Societ},
  \bibinfo{year}{1995}), vol.~\bibinfo{volume}{1}, pp.
  \bibinfo{pages}{311--320}.

\bibitem[{\citenamefont{Starr et~al.}(2002)\citenamefont{Starr, Sastry,
  Douglas, and Glotzer}}]{StarrEtAl02}
\bibinfo{author}{\bibfnamefont{F.}~\bibnamefont{Starr}},
  \bibinfo{author}{\bibfnamefont{S.}~\bibnamefont{Sastry}},
  \bibinfo{author}{\bibfnamefont{J.~F.} \bibnamefont{Douglas}},
  \bibnamefont{and} \bibinfo{author}{\bibfnamefont{S.}~\bibnamefont{Glotzer}},
  \bibinfo{journal}{Phys. Rev. Lett.} \textbf{\bibinfo{volume}{89}},
  \bibinfo{pages}{125501} (\bibinfo{year}{2002}).

\bibitem[{\citenamefont{Bordat et~al.}(2004)\citenamefont{Bordat, Affouard,
  Descamps, and Ngai}}]{BordatNgai04}
\bibinfo{author}{\bibfnamefont{P.}~\bibnamefont{Bordat}},
  \bibinfo{author}{\bibfnamefont{F.}~\bibnamefont{Affouard}},
  \bibinfo{author}{\bibfnamefont{M.}~\bibnamefont{Descamps}}, \bibnamefont{and}
  \bibinfo{author}{\bibfnamefont{K.~L.} \bibnamefont{Ngai}},
  \bibinfo{journal}{Phys. Rev. Lett.} \textbf{\bibinfo{volume}{93}},
  \bibinfo{pages}{105502} (\bibinfo{year}{2004}).

\bibitem[{\citenamefont{Widmer-Cooper and
  Harrowell}(2006{\natexlab{a}})}]{Harrowell06}
\bibinfo{author}{\bibfnamefont{A.}~\bibnamefont{Widmer-Cooper}}
  \bibnamefont{and}
  \bibinfo{author}{\bibfnamefont{P.}~\bibnamefont{Harrowell}},
  \bibinfo{journal}{Phys. Rev. Lett.} \textbf{\bibinfo{volume}{96}},
  \bibinfo{pages}{185701(4)} (\bibinfo{year}{2006}{\natexlab{a}}).

\bibitem[{\citenamefont{Widmer-Cooper et~al.}(2008)\citenamefont{Widmer-Cooper,
  Perry, Harrowell, and Reichman}}]{Harrowell_NP08}
\bibinfo{author}{\bibfnamefont{A.}~\bibnamefont{Widmer-Cooper}},
  \bibinfo{author}{\bibfnamefont{H.}~\bibnamefont{Perry}},
  \bibinfo{author}{\bibfnamefont{P.}~\bibnamefont{Harrowell}},
  \bibnamefont{and} \bibinfo{author}{\bibfnamefont{D.~R.}
  \bibnamefont{Reichman}}, \bibinfo{journal}{Nature Physics}
  \textbf{\bibinfo{volume}{4}}, \bibinfo{pages}{711} (\bibinfo{year}{2008}).

\bibitem[{\citenamefont{Widmer-Cooper et~al.}(2009)\citenamefont{Widmer-Cooper,
  Perry, Harrowell, and Reichman}}]{HarrowellJCP09}
\bibinfo{author}{\bibfnamefont{A.}~\bibnamefont{Widmer-Cooper}},
  \bibinfo{author}{\bibfnamefont{H.}~\bibnamefont{Perry}},
  \bibinfo{author}{\bibfnamefont{P.}~\bibnamefont{Harrowell}},
  \bibnamefont{and} \bibinfo{author}{\bibfnamefont{D.~R.}
  \bibnamefont{Reichman}}, \bibinfo{journal}{J.Chem.Phys.}
  \textbf{\bibinfo{volume}{131}}, \bibinfo{pages}{194508}
  (\bibinfo{year}{2009}).

\bibitem[{\citenamefont{Zhang et~al.}(2009)\citenamefont{Zhang, Srolovitz,
  Douglas, and Warren}}]{DouglasEtAlPNAS2009}
\bibinfo{author}{\bibfnamefont{H.}~\bibnamefont{Zhang}},
  \bibinfo{author}{\bibfnamefont{D.~J.} \bibnamefont{Srolovitz}},
  \bibinfo{author}{\bibfnamefont{J.~F.} \bibnamefont{Douglas}},
  \bibnamefont{and} \bibinfo{author}{\bibfnamefont{J.~A.}
  \bibnamefont{Warren}}, \bibinfo{journal}{Proc. Natl. Acad. Sci. USA}
  \textbf{\bibinfo{volume}{106}}, \bibinfo{pages}{7735} (\bibinfo{year}{2009}).

\bibitem[{\citenamefont{Xia and Wolynes}(2000)}]{XiaWolynes00}
\bibinfo{author}{\bibfnamefont{X.}~\bibnamefont{Xia}} \bibnamefont{and}
  \bibinfo{author}{\bibfnamefont{P.~G.} \bibnamefont{Wolynes}},
  \bibinfo{journal}{PNAS} \textbf{\bibinfo{volume}{97}}, \bibinfo{pages}{2990}
  (\bibinfo{year}{2000}).

\bibitem[{\citenamefont{Larini et~al.}(2008)\citenamefont{Larini, Ottochian,
  De{ }Michele, and Leporini}}]{OurNatPhys}
\bibinfo{author}{\bibfnamefont{L.}~\bibnamefont{Larini}},
  \bibinfo{author}{\bibfnamefont{A.}~\bibnamefont{Ottochian}},
  \bibinfo{author}{\bibfnamefont{C.}~\bibnamefont{De{ }Michele}},
  \bibnamefont{and} \bibinfo{author}{\bibfnamefont{D.}~\bibnamefont{Leporini}},
  \bibinfo{journal}{Nature Physics} \textbf{\bibinfo{volume}{4}},
  \bibinfo{pages}{42} (\bibinfo{year}{2008}).

\bibitem[{\citenamefont{Ottochian et~al.}(2009)\citenamefont{Ottochian, De{
  }Michele, and Leporini}}]{lepoJCP09}
\bibinfo{author}{\bibfnamefont{A.}~\bibnamefont{Ottochian}},
  \bibinfo{author}{\bibfnamefont{C.}~\bibnamefont{De{ }Michele}},
  \bibnamefont{and} \bibinfo{author}{\bibfnamefont{D.}~\bibnamefont{Leporini}},
  \bibinfo{journal}{J. Chem. Phys.} \textbf{\bibinfo{volume}{131}},
  \bibinfo{pages}{224517} (\bibinfo{year}{2009}).

\bibitem[{\citenamefont{Puosi and Leporini}(2011)}]{Puosi11}
\bibinfo{author}{\bibfnamefont{F.}~\bibnamefont{Puosi}} \bibnamefont{and}
  \bibinfo{author}{\bibfnamefont{D.}~\bibnamefont{Leporini}},
  \bibinfo{journal}{J.Phys. Chem. B} \textbf{\bibinfo{volume}{115}},
  \bibinfo{pages}{14046} (\bibinfo{year}{2011}).

\bibitem[{\citenamefont{Puosi et~al.}(2013)\citenamefont{Puosi, Michele, and
  Leporini}}]{SpecialIssueJCP13}
\bibinfo{author}{\bibfnamefont{F.}~\bibnamefont{Puosi}},
  \bibinfo{author}{\bibfnamefont{C.~D.} \bibnamefont{Michele}},
  \bibnamefont{and} \bibinfo{author}{\bibfnamefont{D.}~\bibnamefont{Leporini}},
  \bibinfo{journal}{J. Chem. Phys.} \textbf{\bibinfo{volume}{138}},
  \bibinfo{pages}{12A532} (\bibinfo{year}{2013}).

\bibitem[{\citenamefont{De{ }Michele et~al.}(2011)\citenamefont{De{ }Michele,
  Del{ }Gado, and Leporini}}]{UnivSoftMatter11}
\bibinfo{author}{\bibfnamefont{C.}~\bibnamefont{De{ }Michele}},
  \bibinfo{author}{\bibfnamefont{E.}~\bibnamefont{Del{ }Gado}},
  \bibnamefont{and} \bibinfo{author}{\bibfnamefont{D.}~\bibnamefont{Leporini}},
  \bibinfo{journal}{Soft Matter} \textbf{\bibinfo{volume}{7}},
  \bibinfo{pages}{4025} (\bibinfo{year}{2011}).

\bibitem[{\citenamefont{Ottochian et~al.}(2013)\citenamefont{Ottochian, Puosi,
  Michele, and Leporini}}]{CommentSoftMatter13}
\bibinfo{author}{\bibfnamefont{A.}~\bibnamefont{Ottochian}},
  \bibinfo{author}{\bibfnamefont{F.}~\bibnamefont{Puosi}},
  \bibinfo{author}{\bibfnamefont{C.~D.} \bibnamefont{Michele}},
  \bibnamefont{and} \bibinfo{author}{\bibfnamefont{D.}~\bibnamefont{Leporini}},
  \bibinfo{journal}{Soft Matter} \textbf{\bibinfo{volume}{9}},
  \bibinfo{pages}{7890} (\bibinfo{year}{2013}).

\bibitem[{\citenamefont{Ottochian and
  Leporini}(2011{\natexlab{a}})}]{OttochianLepoJNCS11}
\bibinfo{author}{\bibfnamefont{A.}~\bibnamefont{Ottochian}} \bibnamefont{and}
  \bibinfo{author}{\bibfnamefont{D.}~\bibnamefont{Leporini}},
  \bibinfo{journal}{J. Non-Cryst. Solids} \textbf{\bibinfo{volume}{357}},
  \bibinfo{pages}{298} (\bibinfo{year}{2011}{\natexlab{a}}).

\bibitem[{\citenamefont{Ottochian and
  Leporini}(2011{\natexlab{b}})}]{UnivPhilMag11}
\bibinfo{author}{\bibfnamefont{A.}~\bibnamefont{Ottochian}} \bibnamefont{and}
  \bibinfo{author}{\bibfnamefont{D.}~\bibnamefont{Leporini}},
  \bibinfo{journal}{Phil. Mag.} \textbf{\bibinfo{volume}{91}},
  \bibinfo{pages}{1786} (\bibinfo{year}{2011}{\natexlab{b}}).

\bibitem[{\citenamefont{Puosi and Leporini}(2012{\natexlab{b}})}]{Puosi12SE}
\bibinfo{author}{\bibfnamefont{F.}~\bibnamefont{Puosi}} \bibnamefont{and}
  \bibinfo{author}{\bibfnamefont{D.}~\bibnamefont{Leporini}},
  \bibinfo{journal}{J. Chem. Phys.} \textbf{\bibinfo{volume}{136}},
  \bibinfo{pages}{211101} (\bibinfo{year}{2012}{\natexlab{b}}).

\bibitem[{\citenamefont{Puosi and
  Leporini}(2012{\natexlab{c}})}]{PuosiLepoJCPCor12}
\bibinfo{author}{\bibfnamefont{F.}~\bibnamefont{Puosi}} \bibnamefont{and}
  \bibinfo{author}{\bibfnamefont{D.}~\bibnamefont{Leporini}},
  \bibinfo{journal}{J. Chem. Phys.} \textbf{\bibinfo{volume}{136}},
  \bibinfo{pages}{164901} (\bibinfo{year}{2012}{\natexlab{c}}).

\bibitem[{\citenamefont{Puosi and Leporini}(2013)}]{PuosiLepoJCPCor12_Erratum}
\bibinfo{author}{\bibfnamefont{F.}~\bibnamefont{Puosi}} \bibnamefont{and}
  \bibinfo{author}{\bibfnamefont{D.}~\bibnamefont{Leporini}},
  \bibinfo{journal}{J. Chem. Phys.} \textbf{\bibinfo{volume}{139}},
  \bibinfo{pages}{029901} (\bibinfo{year}{2013}).

\bibitem[{\citenamefont{Buchenau and Zorn}(1992)}]{BuchZorn92}
\bibinfo{author}{\bibfnamefont{U.}~\bibnamefont{Buchenau}} \bibnamefont{and}
  \bibinfo{author}{\bibfnamefont{R.}~\bibnamefont{Zorn}},
  \bibinfo{journal}{Europhys. Lett.} \textbf{\bibinfo{volume}{18}},
  \bibinfo{pages}{523} (\bibinfo{year}{1992}).

\bibitem[{\citenamefont{Andreozzi et~al.}(1998)\citenamefont{Andreozzi,
  Giordano, and Leporini}}]{AndreozziEtAl98}
\bibinfo{author}{\bibfnamefont{L.}~\bibnamefont{Andreozzi}},
  \bibinfo{author}{\bibfnamefont{M.}~\bibnamefont{Giordano}}, \bibnamefont{and}
  \bibinfo{author}{\bibfnamefont{D.}~\bibnamefont{Leporini}},
  \bibinfo{journal}{J. Non-Cryst. Solids} \textbf{\bibinfo{volume}{235}},
  \bibinfo{pages}{219} (\bibinfo{year}{1998}).

\bibitem[{\citenamefont{Scopigno et~al.}(2003)\citenamefont{Scopigno, Ruocco,
  Sette, and Monaco}}]{ScopignoEtAl03}
\bibinfo{author}{\bibfnamefont{T.}~\bibnamefont{Scopigno}},
  \bibinfo{author}{\bibfnamefont{G.}~\bibnamefont{Ruocco}},
  \bibinfo{author}{\bibfnamefont{F.}~\bibnamefont{Sette}}, \bibnamefont{and}
  \bibinfo{author}{\bibfnamefont{G.}~\bibnamefont{Monaco}},
  \bibinfo{journal}{Science} \textbf{\bibinfo{volume}{302}},
  \bibinfo{pages}{849} (\bibinfo{year}{2003}).

\bibitem[{\citenamefont{Sokolov et~al.}(1993)\citenamefont{Sokolov, R\"ossler,
  Kisliuk, and Quitmann}}]{SokolPRL}
\bibinfo{author}{\bibfnamefont{A.~P.} \bibnamefont{Sokolov}},
  \bibinfo{author}{\bibfnamefont{E.}~\bibnamefont{R\"ossler}},
  \bibinfo{author}{\bibfnamefont{A.}~\bibnamefont{Kisliuk}}, \bibnamefont{and}
  \bibinfo{author}{\bibfnamefont{D.}~\bibnamefont{Quitmann}},
  \bibinfo{journal}{Phys. Rev. Lett.} \textbf{\bibinfo{volume}{71}},
  \bibinfo{pages}{2062} (\bibinfo{year}{1993}).

\bibitem[{\citenamefont{Buchenau and Wischnewski}(2004)}]{Buchenau04}
\bibinfo{author}{\bibfnamefont{U.}~\bibnamefont{Buchenau}} \bibnamefont{and}
  \bibinfo{author}{\bibfnamefont{A.}~\bibnamefont{Wischnewski}},
  \bibinfo{journal}{Phys. Rev. B} \textbf{\bibinfo{volume}{70}},
  \bibinfo{pages}{092201} (\bibinfo{year}{2004}).

\bibitem[{\citenamefont{Novikov and Sokolov}(2004)}]{NoviSoko04}
\bibinfo{author}{\bibfnamefont{V.~N.} \bibnamefont{Novikov}} \bibnamefont{and}
  \bibinfo{author}{\bibfnamefont{A.~P.} \bibnamefont{Sokolov}},
  \bibinfo{journal}{Nature} \textbf{\bibinfo{volume}{431}},
  \bibinfo{pages}{961} (\bibinfo{year}{2004}).

\bibitem[{\citenamefont{Novikov et~al.}(2005)\citenamefont{Novikov, Ding, and
  Sokolov}}]{NovikovEtAl05}
\bibinfo{author}{\bibfnamefont{V.~N.} \bibnamefont{Novikov}},
  \bibinfo{author}{\bibfnamefont{Y.}~\bibnamefont{Ding}}, \bibnamefont{and}
  \bibinfo{author}{\bibfnamefont{A.~P.} \bibnamefont{Sokolov}},
  \bibinfo{journal}{Phys.Rev.E} \textbf{\bibinfo{volume}{71}},
  \bibinfo{pages}{061501} (\bibinfo{year}{2005}).

\bibitem[{\citenamefont{Yannopoulos and Johari}(2006)}]{Johari06}
\bibinfo{author}{\bibfnamefont{S.~N.} \bibnamefont{Yannopoulos}}
  \bibnamefont{and} \bibinfo{author}{\bibfnamefont{G.~P.}
  \bibnamefont{Johari}}, \bibinfo{journal}{Nature}
  \textbf{\bibinfo{volume}{442}}, \bibinfo{pages}{E7} (\bibinfo{year}{2006}).

\bibitem[{\citenamefont{Egelstaff}(1992)}]{Egelstaff:1992fk}
\bibinfo{author}{\bibfnamefont{P.~A.} \bibnamefont{Egelstaff}},
  \emph{\bibinfo{title}{An introduction to the liquid state}}
  (\bibinfo{publisher}{Clarendon Press}, \bibinfo{address}{Oxford},
  \bibinfo{year}{1992}), \bibinfo{edition}{2nd} ed.

\bibitem[{\citenamefont{Hansen and McDonald}(2006)}]{HansenMcDonaldIIIEd}
\bibinfo{author}{\bibfnamefont{J.~P.} \bibnamefont{Hansen}} \bibnamefont{and}
  \bibinfo{author}{\bibfnamefont{I.~R.} \bibnamefont{McDonald}},
  \emph{\bibinfo{title}{Theory of Simple Liquids, 3rd Ed.}}
  (\bibinfo{publisher}{Academic Press}, \bibinfo{year}{2006}).

\bibitem[{not()}]{note3}
\bibinfo{howpublished}{Note that $t^\star$ is one order of magnitude shorter
  than the times considered in ref. \cite{Harrowell06}.}

\bibitem[{\citenamefont{Jedlovszky}(1999)}]{JedloJCP}
\bibinfo{author}{\bibfnamefont{P.}~\bibnamefont{Jedlovszky}},
  \bibinfo{journal}{J. Chem. Phys.} \textbf{\bibinfo{volume}{111}},
  \bibinfo{pages}{5975} (\bibinfo{year}{1999}).

\bibitem[{\citenamefont{Wikfeldt et~al.}(2010)\citenamefont{Wikfeldt, Leetmaa,
  Mace, Nilsson, and Pettersson}}]{Wikfeldt10}
\bibinfo{author}{\bibfnamefont{K.~T.} \bibnamefont{Wikfeldt}},
  \bibinfo{author}{\bibfnamefont{M.}~\bibnamefont{Leetmaa}},
  \bibinfo{author}{\bibfnamefont{A.}~\bibnamefont{Mace}},
  \bibinfo{author}{\bibfnamefont{A.}~\bibnamefont{Nilsson}}, \bibnamefont{and}
  \bibinfo{author}{\bibfnamefont{L.~G.~M.} \bibnamefont{Pettersson}},
  \bibinfo{journal}{J. Chem. Phys.} \textbf{\bibinfo{volume}{132}},
  \bibinfo{pages}{104513} (\bibinfo{year}{2010}).

\bibitem[{\citenamefont{Stirnemann and Laage}(2012)}]{Stirnemann12}
\bibinfo{author}{\bibfnamefont{G.}~\bibnamefont{Stirnemann}} \bibnamefont{and}
  \bibinfo{author}{\bibfnamefont{D.}~\bibnamefont{Laage}}, \bibinfo{journal}{J.
  Chem. Phys.} \textbf{\bibinfo{volume}{137}}, \bibinfo{pages}{031101}
  (\bibinfo{year}{2012}).

\bibitem[{\citenamefont{Ruocco et~al.}(1992)\citenamefont{Ruocco, Sampoli, and
  Vallauri}}]{H2SJCP}
\bibinfo{author}{\bibfnamefont{G.}~\bibnamefont{Ruocco}},
  \bibinfo{author}{\bibfnamefont{M.}~\bibnamefont{Sampoli}}, \bibnamefont{and}
  \bibinfo{author}{\bibfnamefont{R.}~\bibnamefont{Vallauri}},
  \bibinfo{journal}{J. Chem. Phys.} \textbf{\bibinfo{volume}{96}},
  \bibinfo{pages}{6167} (\bibinfo{year}{1992}).

\bibitem[{\citenamefont{Krekelberg et~al.}(2006)\citenamefont{Krekelberg,
  Ganesan, and Truskett}}]{Krekelberg06}
\bibinfo{author}{\bibfnamefont{W.~P.} \bibnamefont{Krekelberg}},
  \bibinfo{author}{\bibfnamefont{V.}~\bibnamefont{Ganesan}}, \bibnamefont{and}
  \bibinfo{author}{\bibfnamefont{T.~M.} \bibnamefont{Truskett}},
  \bibinfo{journal}{J. Chem. Phys.} \textbf{\bibinfo{volume}{124}},
  \bibinfo{pages}{214502} (\bibinfo{year}{2006}).

\bibitem[{\citenamefont{Sega et~al.}(2004)\citenamefont{Sega, Jedlovszky,
  Medvedev, and Vallauri}}]{SegaJCP}
\bibinfo{author}{\bibfnamefont{M.}~\bibnamefont{Sega}},
  \bibinfo{author}{\bibfnamefont{P.}~\bibnamefont{Jedlovszky}},
  \bibinfo{author}{\bibfnamefont{N.~N.} \bibnamefont{Medvedev}},
  \bibnamefont{and} \bibinfo{author}{\bibfnamefont{R.}~\bibnamefont{Vallauri}},
  \bibinfo{journal}{J. Chem. Phys.} \textbf{\bibinfo{volume}{121}},
  \bibinfo{pages}{2422} (\bibinfo{year}{2004}).

\bibitem[{\citenamefont{D.Rigby and Roe}(1990)}]{Rigbyt}
\bibinfo{author}{\bibnamefont{D.Rigby}} \bibnamefont{and}
  \bibinfo{author}{\bibfnamefont{R.~J.} \bibnamefont{Roe}},
  \bibinfo{journal}{Macromolecules} \textbf{\bibinfo{volume}{23}},
  \bibinfo{pages}{5312} (\bibinfo{year}{1990}).

\bibitem[{\citenamefont{Grest and Kremer}(1986)}]{GrestPRA33}
\bibinfo{author}{\bibfnamefont{G.~S.} \bibnamefont{Grest}} \bibnamefont{and}
  \bibinfo{author}{\bibfnamefont{K.}~\bibnamefont{Kremer}},
  \bibinfo{journal}{Phys. Rev. A} \textbf{\bibinfo{volume}{33}},
  \bibinfo{pages}{3628} (\bibinfo{year}{1986}).

\bibitem[{\citenamefont{Andersen}(1980)}]{Andersen80}
\bibinfo{author}{\bibfnamefont{H.~C.} \bibnamefont{Andersen}},
  \bibinfo{journal}{J. Chem. Phys.} \textbf{\bibinfo{volume}{72}},
  \bibinfo{pages}{2384} (\bibinfo{year}{1980}).

\bibitem[{\citenamefont{Nos\'e}(1984)}]{NTVnose}
\bibinfo{author}{\bibfnamefont{S.}~\bibnamefont{Nos\'e}}, \bibinfo{journal}{J.
  Chem. Phys.} \textbf{\bibinfo{volume}{81}}, \bibinfo{pages}{511}
  (\bibinfo{year}{1984}).

\bibitem[{\citenamefont{Tuckerman et~al.}(1992)\citenamefont{Tuckerman, Berne,
  and Martyna}}]{respa}
\bibinfo{author}{\bibfnamefont{M.~E.} \bibnamefont{Tuckerman}},
  \bibinfo{author}{\bibfnamefont{B.~J.} \bibnamefont{Berne}}, \bibnamefont{and}
  \bibinfo{author}{\bibfnamefont{G.~J.} \bibnamefont{Martyna}},
  \bibinfo{journal}{J. Chem. Phys.} \textbf{\bibinfo{volume}{97}},
  \bibinfo{pages}{1990} (\bibinfo{year}{1992}).

\bibitem[{\citenamefont{Bueche}(1962)}]{Bueche}
\bibinfo{author}{\bibfnamefont{F.}~\bibnamefont{Bueche}},
  \emph{\bibinfo{title}{Physical Properties of Polymers}}
  (\bibinfo{publisher}{Interscience, New York}, \bibinfo{year}{1962}).

\bibitem[{\citenamefont{Barbieri
  et~al.}(2004{\natexlab{b}})\citenamefont{Barbieri, Prevosto, Lucchesi, and
  Leporini}}]{BarbieriEtAl2004}
\bibinfo{author}{\bibfnamefont{A.}~\bibnamefont{Barbieri}},
  \bibinfo{author}{\bibfnamefont{D.}~\bibnamefont{Prevosto}},
  \bibinfo{author}{\bibfnamefont{M.}~\bibnamefont{Lucchesi}}, \bibnamefont{and}
  \bibinfo{author}{\bibfnamefont{D.}~\bibnamefont{Leporini}},
  \bibinfo{journal}{J. Phys.: Condens. Matter} \textbf{\bibinfo{volume}{16}},
  \bibinfo{pages}{6609} (\bibinfo{year}{2004}{\natexlab{b}}).

\bibitem[{\citenamefont{Widmer-Cooper and
  Harrowell}(2006{\natexlab{b}})}]{HarrowellFreeVolume06}
\bibinfo{author}{\bibfnamefont{A.}~\bibnamefont{Widmer-Cooper}}
  \bibnamefont{and}
  \bibinfo{author}{\bibfnamefont{P.}~\bibnamefont{Harrowell}},
  \bibinfo{journal}{J. Non-Cryst Solids} \textbf{\bibinfo{volume}{352}},
  \bibinfo{pages}{5098} (\bibinfo{year}{2006}{\natexlab{b}}).

\bibitem[{\citenamefont{Berthier and Jack}(2007)}]{BerthierJackPRE07}
\bibinfo{author}{\bibfnamefont{L.}~\bibnamefont{Berthier}} \bibnamefont{and}
  \bibinfo{author}{\bibfnamefont{R.~L.} \bibnamefont{Jack}},
  \bibinfo{journal}{Phys. Rev. E} \textbf{\bibinfo{volume}{76}},
  \bibinfo{pages}{041509} (\bibinfo{year}{2007}).

\bibitem[{\citenamefont{F.Puosi and D.Leporini}()}]{Elastico2}
\bibinfo{author}{\bibnamefont{F.Puosi}} \bibnamefont{and}
  \bibinfo{author}{\bibnamefont{D.Leporini}},
  \bibinfo{howpublished}{arXiv:1108.4629}.

\end{thebibliography}
\end{document}